\documentstyle[twocolumn]{mn2e}

\def\be{\begin{equation}}
\def\ee{\end{equation}}
\def\beq{\begin{eqnarray}}
\def\eeq{\end{eqnarray}}
\def\GAIA{{\it GAIA}}
\def\Hipparcos{{\it Hipparcos}}
\def\kmsMpc{ {\rm kms}^{-1}{\rm Mpc}^{-1}}
\def\Lsun{ {L_\odot}}

\def\f{f_{\rm II}}
\def\spose#1{\hbox to 0pt{#1\hss}}
\def\lta{\mathrel{\spose{\lower 3pt\hbox{$\sim$}} \raise
2.0pt\hbox{$<$}}}
\def\gta{\mathrel{\spose{\lower 3pt\hbox{$\sim$}} \raise
2.0pt\hbox{$>$}}}

\input epsf

\title[Supernovae with ``Super-\Hipparcos'']
      {Supernovae with ``Super-\Hipparcos''}

\author[V.A. Belokurov \& N.W. Evans]
        {V.A. Belokurov$^1$ \& N.W. Evans$^{1,2}$ \\
        $^1$ Theoretical Physics, 1 Keble Rd, Oxford, OX1 3NP \\
        $^2$ Institute of Astronomy, Madingley Rd, Cambridge, CB3 0HA}

\date{}
\begin{document}
\maketitle
\label{firstpage}

\begin{abstract}
\GAIA\ is the ``super-\Hipparcos'' satellite scheduled for launch in
2010 by the European Space Agency. It is a scanning satellite that
carries out multi-colour, multi-epoch photometry on all objects
brighter than 20th magnitude.  We conduct detailed simulations of
supernovae (SNe) detection by \GAIA.  Supernovae of each type are
chosen according to the observed distributions of absolute magnitudes,
and located in nearby galaxies according to the local large-scale
structure. Using an extinction model of the Galaxy and the scanning
law of the \GAIA\ satellite, we calculate how many SNe are detectable
as a function of the phase of the lightcurve.  Our study shows that
\GAIA\ will report data on $\sim 21\,400$ SNe during the five-year
mission lifetime, of which $\sim 14\,300$ are SNe Ia, $\sim 1400$ are
SNe Ib/c and $\sim 5700$ are SNe II. Using the simulations, we
estimate that the numbers caught before maximum are $\sim 6300$ SNe
Ia, $\sim 500$ SNe Ib/c and $\sim 1700$ SNe II. During the mission
lifetime, \GAIA\ will issue about $5$ SNe alerts a day.

The most distant SNe accessible to \GAIA\ are at a redshift $z \sim
0.14$ and so \GAIA\ will provide a huge sample of local SNe.  There
will be many examples of the rarer subluminous events, over-luminous
events, SNe Ib/c and SNe II-L. SNe rates will be found as a function
of galaxy type, as well as extinction and position in the host galaxy.
Amongst other applications, there may be about 26 SNe each year for
which detection of gravitational waves is possible and about 180 SNe
each year for which detection of gamma-rays is possible.  \GAIA's
astrometry will provide the SN position to better than
milliarcseconds, offering opportunities for the identification of
progenitors in nearby galaxies and for studying the spatial
distribution of SNe of different types in galaxies.
\end{abstract}

\begin{keywords}
supernovae: general -- neutrinos -- gamma rays: theory --
gravitational waves -- gravitational lensing
\end{keywords}

\section{Introduction}

\GAIA\ is the super-\Hipparcos\ satellite that is planned for launch
in about 2010 (see ``http//astro.estec.esa.nl/gaia'' or Perryman et
al. 2001). It is the successor to the pioneering \Hipparcos\ satellite
which revolutionised our knowledge of the solar neighborhood. \GAIA\
gathers 4 colour broad band and 11 colour narrow band photometry on
all objects brighter than $20$th magnitude and it performs
spectroscopy in the range 850-875 nm on all objects brighter than
$18$th magnitude.  Here, we discuss the capabilities of this
remarkable satellite for supernovae (SNe) detection.

SNe are divided into type I and II primarily on the basis of their
spectra, with lightcurve shape as a secondary diagnostic.  SNe Ia are
thermonuclear explosions in white dwarf stars which have accreted too
much matter from a companion. It is unclear whether the companion is
also a white dwarf or is a main sequence/red giant star. In fact,
whether the progenitors of SNe Ia are doubly degenerate or singly
degenerate binaries is one of the major unsolved problems in the
subject.  SNe Ib/c and II originate in the core collapse of massive
stars. SNe Ia have found ready application in cosmology. Their
intrinsic brightness means that they can be detected to enormous
distances.  Although their peak luminosities vary by a factor of 10,
Phillips (1993) found a correlation between the peak absolute
magnitude and the rate of decline, which enables SNe Ia to be
calibrated and used as ``standard candles''.  Claims for an
accelerating universe (e.g., Riess et al. 1998, Schmidt et al. 1998,
Perlmutter et al. 1999) partly rest on their use as distance
estimators, yet there are obvious concerns regarding systematic errors
(e.g., Leibundgut 2000, 2001). So, a major thrust of modern SNe
studies is to understand and quantify the differences in the
morphology of SNe Ia lightcurves and the scatter in the Phillips
relation.

SNe surveys go back at least as far as Zwicky (1938). Amongst the most
influential are the Asiago SNe survey (Ciatti \& Rosino 1978) and the
Cal\'an/Tololo SNe survey (Hamuy et al. 1993).  At the moment, there
is some, albeit limited, information on $\sim 2000$ SNe available in
the standard catalogues (see Tsvetkov et al.  1998 or
``http://www.sai.msu.su/sn/sncat/'' and Barbon et al. 1999 or
``http://web.pd.astro.it/supern/''). Perhaps only as many as $\sim
300$ SNe have reasonably detailed lightcurves (e.g., Leibundgut et
al. 1991, Hamuy et al. 1996, Leibundgut 2000). Even with the limited
dataset available, there is a considerable spread in the intrinsic
properties and uncertainty as to the rate, type and location of SNe as
a function of host galaxy.

\GAIA\ is an ideal tool to study nearby SNe (within a few hundred
Mpc). \GAIA\ will provide a huge dataset of high quality local SNe Ia
in which any deviations from ``standard candles'' can be analysed.  As
the dataset is so large, there will be good numbers of rarer phenomena,
such as subluminous SNe and SNe Ib/c.  Earlier papers (H{\o}g,
Fabricius \& Makarov 1999; Tammann \& Reindl 2002) have already
provided rough estimates of the numbers of SNe Ia that GAIA will
discover. Here, we carry out detailed simulations using a Galactic
extinction model and the satellite scanning law to compute the numbers
of SNe detected as a function of the phase of the lightcurve. We show
that \GAIA\ will record data on at least $21\, 400$ SNe during the
five-year mission lifetime. This breaks down into $\sim 14\, 300$ SNe
1a, $\sim 1400$ SNe Ib/c and $\sim 5700$ SNe II. These SNe span a
redshift range up to $z \sim 0.14$.  \GAIA\ will probably alert on all
SNe detected before maximum. These numbers are $\sim 6300$ SNe Ia,
$\sim 500$ SNe Ib/c and $\sim 1700$ SNe II during the whole
mission. In other words, \GAIA\ will issue $\sim 1700$ SNe alerts a
year or $\sim 5$ alerts a day. Roughly $75 \%$ of all alerts will be
SNe Ia, the remainder will be SNe Ib/c and II.  All these numbers are
lower limits and may be increased by at a factor of $\sim 2$ depending
on the SNe contribution from low-luminosity galaxies.

\begin{table*}
\begin{center}
\begin{tabular}{l|ccccc}\hline
 & E-S0 & S0/a, Sa & Sab, Sb & Sbc-Sd & Sdm-Im\\
\hline
Type Ia& 0.42 & 0.21 & 0.21  & 0.21  & 0.21\\
Type Ib/c & - & 0.01 & 0.11 & 0.33 & 0.38 \\
Type II& - & 0.07 & 0.57 & 1.66 & 1.78 \\
\hline
\end{tabular}
\end{center}
\caption{SNe explosion rates (in units of SNe per $10^{10} \Lsun$ per
century) according to Hubble type, adapted from the van den Bergh \&
Tammann (1991) values using a Hubble constant of $65\ \kmsMpc$. The
galaxy types follow the classification scheme of de Vaucouleurs (1959;
see also van den Bergh 1998).}
\label{table:rates}
\end{table*}
\begin{table*}
\begin{center}
\begin{tabular}{l|ccccc}\hline
 & E-S0 & S0/a, Sa & Sab, Sb & Sbc-Sd & Sdm-Im\\
\hline
N (galaxies) & 1200 (4850) & 679 (3024) & 867 (3694) & 1993 (6957) & 
1798 (3701) \\
Type Ia& 29 (80) & 7 (11) & 11 (31) & 19 (52) & 16 (16) \\
Type Ib/c & 0 (0) & 1 (1) & 6 (16) & 30 (80) & 29 (29) \\
Type II& 0 (0) & 2 (4) & 31 (85) & 152 (412) & 135 (135) \\
\hline
\end{tabular}
\end{center}
\caption{Number of galaxies of each Hubble type within 75 Mpc in the
CfA catalogue.  Also shown is the number of SNe within 75 Mpc during
the 5 year mission lifetime.  For comparison, the figures in
parentheses are based on an extrapolation of the galaxy LF that is
flat down to an absolute magnitude of $-14$.}
\label{table:cfa}
\end{table*}
\begin{figure}
\epsfxsize=10cm \centerline{\epsfbox{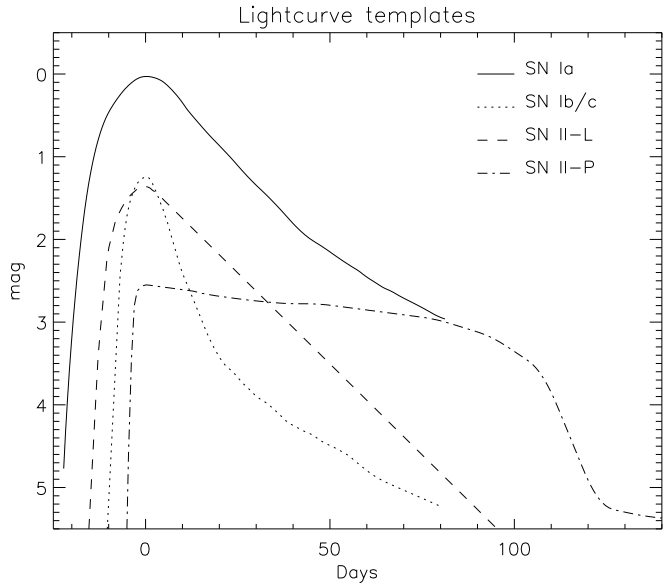}}
\caption{This shows the templates adopted for SNe Ia, Ib/c, II-L and
II-P in the Monte Carlo simulations. The templates come from data on
SN 1991T by Lira et al. (1998), on SN 1994I by Richmond et al. (1996),
on SN 1990K by Cappellaro et al. (1995) and SN 1999em by Hamuy et
al. (2001). The graphs show relative magnitude versus time in days.}
\label{fig:templates}
\end{figure}

\section{Monte Carlo Simulations}

\GAIA's G band is very close to the conventional V band for a wide
colour range (ESA 2000). Richardson et al. (2002) use the Asiago
Supernova Catalogue to study the V band absolute magnitude
distributions according to type.  They find that the mean absolute
magnitude of SNe Ia at maximum is $-18.99$. This figure includes the
contribution from the internal absorption of the host galaxy, as well
as a correction to our preferred value of the Hubble constant of $65
\kmsMpc$, which we use henceforth.  As \GAIA's limiting magnitude is
$G \sim 20$, this means that the most distant SNe Ia accessible to
\GAIA\ are $\sim 630$ Mpc away. Similarly, the mean absolute magnitude
of SNe Ib/c at maximum is $-17.75$, so that the most distant SNe Ib/c
detectable by \GAIA\ are $\sim 355$ Mpc away. SNe II are subdivided
further according to lightcurve into linear (L) and plateau (P) types.
The mean absolute magnitude of II-L type is $-17.63$ and of II-P type
is $-16.44$ (Richardson et al. 2002).  These correspond to distances
of $\sim 335$ Mpc and $\sim 195$ Mpc respectively.  Such distances
emphasise that \GAIA\ is the ideal tool to discover relatively nearby
SNe, but will not make any contribution to the searches for high
redshift SNe.

The rates with which SNe occur in different galaxies at low redshift
are given in Table~\ref{table:rates}, inferred from van den Bergh \&
Tammann (1991).  This gives the number of SNe per century per $10^{10}
\Lsun$. Although the data come with substantial uncertainties, there
are three recent studies that provide some supporting evidence. First,
the EROS collaboration (Hardin et al. 2000) found a SNe Ia rate of
$\sim 0.18$ per century per $10^{10} \Lsun$ for redshifts $z$ in the
range $0.02$ to $0.2$. Second, Pain et al. (1996) found a SNe Ia rate
of $\sim 0.36$ per century per $10^{10} \Lsun$ at the somewhat higher
redshift of $z \sim 0.4$. Third, the rates in Cappellaro et al.
(1997) are of the same order as van den Bergh \& Tammann (1991),
although typically lower by a factor of two.

The host galaxies follow the local large-scale structure.  We use the
latest version of the CfA redshift catalogue (Huchra et al. 1992, see
``http://cfa-www.harvard.edu/\~{}huchra/zcat/''). This contains the
sky positions and heliocentric velocities of $\sim 20\,000$
galaxies. On plotting numbers of galaxies versus distance, the graph
peaks at $\sim 75$ Mpc and thence shows a steady decline. This
suggests that the catalogue can be used out to at most $\sim 75$ Mpc.
The number of galaxies of each Hubble type in the CfA catalogue within
75 Mpc is listed in Table~\ref{table:cfa}. These numbers must be
regarded as lower limits to the true numbers within 75 Mpc, as the
luminosity functions (LFs) derived from the CfA catalogue are
incomplete at faint magnitudes.  To take this into account, we assume
that after reaching maximum the galaxy LF remains flat down to an
absolute magnitude of $-14$.  Given in parentheses in
Table~\ref{table:cfa} are the number of galaxies and SNe within 75 Mpc
assuming such a flat LF. These numbers can be regarded as upper
limits.  Beyond 75 Mpc, we assume that the distribution of galaxies is
homogeneous and that the number scales like $D^3$ where $D$ is the
heliocentric distance.  Given the numbers of galaxies and the observed
rates, we can then straightforwardly compute the total number of SNe
that explode during the five-year \GAIA\ mission lifetime and are
brighter than the limiting magnitude.  In all, there are at least
$\sim 48\,000$ SNe Ia and $\sim 7000$ SNe Ib/c. The numbers for SNe II
depend on the relative frequency of II-L with respect to II-P, which
is not very well-known. Henceforth, we denote the fraction of all SNe
II that are L-type by $\f$. Then, the numbers of SNe II that explode
are $\sim 28\,500 \f + 5600 (1-\f)$.  These numbers are lower limits
for two reasons -- first because no contribution from faint galaxies
is included and second because \GAIA's limiting magnitude may be
deeper than 20th in practice. Allowing for faint galaxies with the
flat LF gives results a factor $\sim 2$ times larger.
\begin{figure}
\epsfxsize=10cm \centerline{\epsfbox{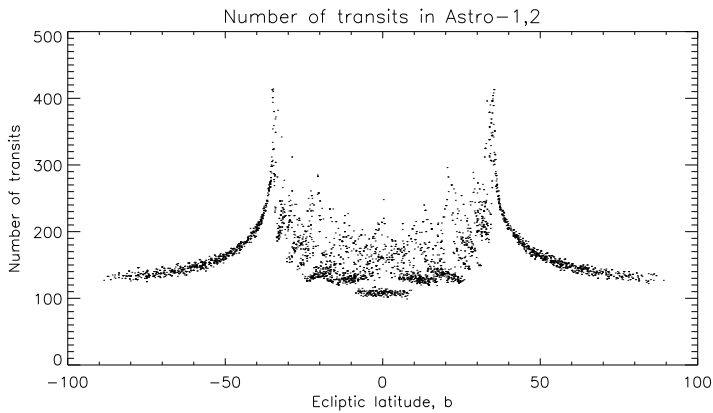}}
\caption{This shows the number of transits made by the ASTRO-1 and
ASTRO-2 telescopes as a function of ecliptic latitude. The figure is
drawn by choosing 5000 random directions on the sky, and computing the
ecliptic latitude and number of transits using software supplied by
L. Lindegren. The transits are strongly clustered into groups of
between two and five.}
\label{fig:transits}
\end{figure}

To assess the efficiency of SN detection, we perform Monte Carlo
simulations (e.g., Li, Filippenko \& Riess 2001).  First, the distance
to the host galaxy is picked from the $D^3$ distribution. If $D < 75$
Mpc, then the galaxy is chosen from the CfA catalogue. If $D > 75$
Mpc, then the galaxy positions are distributed uniformly over the sky.
Next, we choose the SN absolute magnitudes from Gaussian distributions
with means and dispersions given in Table~\ref{table:absmag}. These
numbers already includes the effects of absorption in the host galaxy.
The contribution to dimming of the SN from Galactic absorption is
calculated using software kindly supplied by R. Drimmel, which is
based on the extinction model of Drimmel \& Spergel (2001).

The lightcurve can now be generated using the templates illustrated in
Figure~\ref{fig:templates}. For SNe Ia, we use the data of Lira et
al. (1998) on SN 1991T as a standard template. This is an unusually
bright SN Ia with a peculiar spectrum (Saha et al. 2001). However, our
algorithm chooses the magnitude at maximum from a Gaussian
distribution and so only the shape of the template is important.
Therefore, although SN 1991T is overluminous, this does not cause
exaggeration of our results.  We reconstruct the rising part of the
lightcurve of SNe Ia using the fact that luminosity is proportional to
the square of the time since explosion (e.g., Riess et al. 1999).  For
SNe Ib/c, we use the photometry on SN 1994I from Richmond et
al. (1996). This is a SN Ic, but we use it as a template for all SN
Ib/c, taking the view that there is little point in trying to
distinguish between SN Ibs and SN Ics in our simulations. It is not
even clear that they are intrinsically very different, other than in
the location of the ionizing source for helium.  There is considerable
diversity in SNe II lightcurves, as illustrated in Patat et
al. (1994).  We use the photometry of Cappellaro et al. (1995) on SN
1990K as the basis of our standard template for SNe II-L. However,
this is missing the rising part, which is reconstructed using the mean
lightcurve data in Doggett \& Branch (1985).  We use the photometry of
Hamuy et al. (2001) on SN 1999em as a template for SN II-P.

Lastly, the lightcurves are sampled according to \GAIA's scanning law
using software kindly supplied by L. Lindegren.  \GAIA\ has three
telescopes on board.  ASTRO-1 and ASTRO-2 perform astrometry and
broad-band photometry on all objects brighter than $G \sim 20$.
SPECTRO performs spectroscopy and medium-band photometry. The
spacecraft is assumed to rotate about its spin axis once every $\sim
3$ hours, performing a great circle scan. (This corresponds to the
prototype, although a slower spin rate may be adopted in the final
mission).  The spin axis is constrained to move on a Sun-centered cone
of $55^\circ$ with a period of 76 days, forcing the plane of the scan
to sway back and forth. The axis of the cone follows the yearly solar
motion (see ESA 2000). The number of times an object enters the fields
of view of ASTRO-1 and ASTRO-2 during mission lifetime depends on
ecliptic latitude and is shown in Figure~\ref{fig:transits} (see e.g.,
Belokurov \& Evans 2002).

\begin{table}
\begin{center}
\begin{tabular}{l|cc}\hline
SN Type & $M_G$ & $\sigma_G$ \\
\hline
Type Ia   & -18.99  & 0.76 \\
Type Ib/c & -17.75  & 1.29 \\
Type II-L & -17.63   & 0.88 \\
Type II-P & -16.44   & 1.23 \\
\hline
\end{tabular}
\end{center}
\caption{The means and dispersions in SNe absolute magnitude at
maximum adopted for the Monte Carlo simulations. The numbers are
derived from the uncorrected distributions in Richardson et
al. (2002), but are adjusted to our preferred Hubble constant of $65\
\kmsMpc$.}
\label{table:absmag}
\end{table}
\begin{figure}
\begin{center}
\epsfxsize=10cm \centerline{\epsfbox{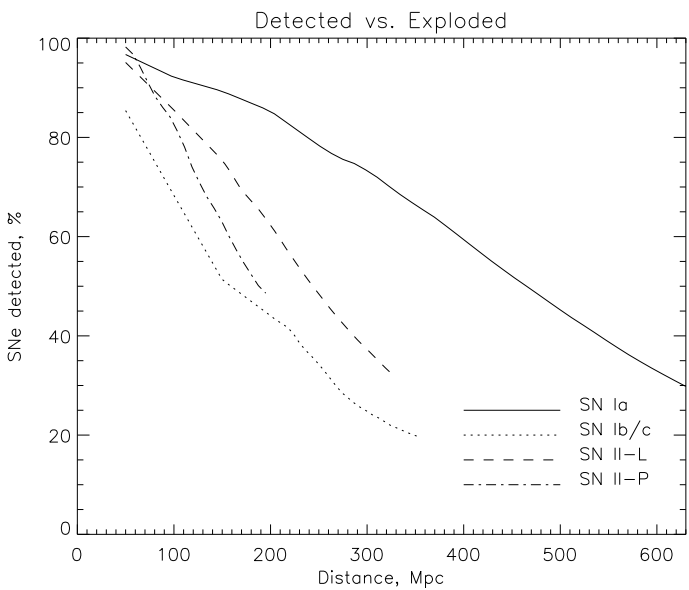}}
\end{center}
\caption{This shows the total number of SNe detected within distance
$D$ as a fraction of the total number exploded. Within 630 Mpc, \GAIA\
detects $\sim 30 \%$ of all SNe Ia. Within 355 Mpc, \GAIA\ detects
$\sim 20 \%$ of all SNe Ib/c. For SNe II-L, \GAIA\ detects $\sim 31
\%$ within 335 Mpc. Finally, for SNe II-P, \GAIA\ detects $\sim 48 \%$
within 195 Mpc. Note that by detection, we mean that \GAIA\ records at
least one datapoint on the standard SNe templates shown in
Figure~\ref{fig:templates}.}
\label{fig:detect}
\end{figure}

\section{Results}

\subsection{Numbers Detected}

Figure~\ref{fig:detect} shows the fraction of SNe within a distance
$D$ which enter the field of view of \GAIA's telescopes ASTRO-1 and
ASTRO-2.  \GAIA\ records data on $30 \%$ of all the SNe Ia within 630
Mpc, which marks the limit of the most distant SNe Ia accessible.  The
distance cut-offs for the intrinsically less bright SNe Ib/c, II-L and
II-P are 355 Mpc, 335 Mpc and 195 Mpc respectively. \GAIA\ records
data on $\sim 20 \%$ of all SNe Ib/c, $\sim 31 \%$ of all SNe II-L and
$\sim 48 \%$ SNe II-P within these distances.  This means that \GAIA\
will provide some (perhaps rather limited) information on $14\,300$
SNe Ia and $1400$ SNe Ib/c during its five-year mission.  For SNe II,
the number depends on the relative frequency $\f$ and is $\sim 8700 \f
+ 2700 (1-\f)$.  If SNe II-L and SNe II-P occur equally frequently
($\f = 0.5$), then the total number of SNe II is $\sim 5700$. In other
words, \GAIA\ will provide some information on $\sim 21\,400$ SNe in
total. For comparison, H{\o}g et al. (1999) used simple scaling
arguments to estimate that the total number of SNe in the \GAIA\
observations would be $\sim 100\,000$.

These are huge numbers, both compared to the sizes of existing
catalogues and to the likely datasets gathered by other planned space
missions.  Almost all the SNe that \GAIA\ misses explode in the 20
days just after \GAIA\ samples that location in the sky. Before the
next transit of ASTRO-1 or ASTRO-2, the SN reaches maximum and then
fades to below \GAIA's limiting magnitude ($G \sim 20$).  It may be
wondered whether some SNe are missed because light from the background
galaxy can overwhelm the SN.  This is clearly a problem for distant
galaxies, which are wholly contained within \GAIA's PSF ($\sim .35''$
at FWHM). However, rough calculations show that this is not a problem
for SNe Ib/c and II, as they occur relatively close by; we estimate
that it may affect $\lta 10 \%$ of SNe Ia.

\begin{figure*}
\begin{center}
\epsfxsize=14cm \centerline{\epsfbox{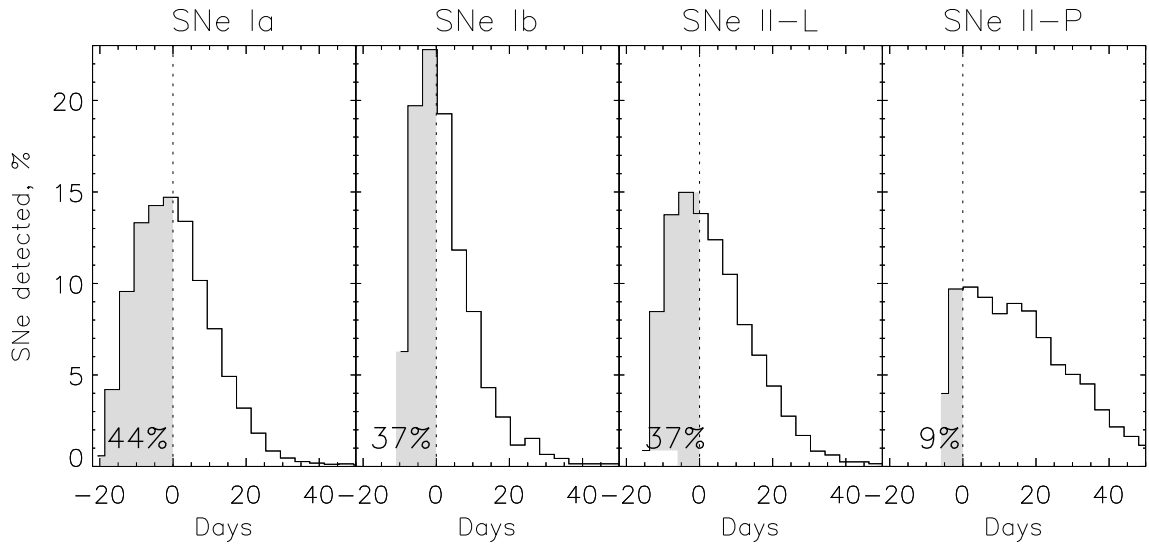}}
\end{center}
\vskip0.5truecm
\caption{This shows histograms of the numbers of detected SNe against
phase of the lightcurve. The shaded area corresponds to the fraction
of SNe caught before the maximum of the lightcurve.}
\label{fig:phase}
\end{figure*}

Figure~\ref{fig:phase} shows the fraction of the detected SNe as a
function of phase of the lightcurve. Some $44 \%$ of the detected SNe
Ia are caught before maximum, $37 \%$ of the detected SNe Ib/c, $37
\%$ of the detected SNe II-L and $9 \%$ of the detected SNe II-P. The
low fraction for SNe II-P is largely a consequence of the fact that
they are intrinsically the faintest.  The numbers of each type of SNe
caught before maximum by \GAIA\ are recorded in
Table~\ref{table:nosne}. The total number of all SNe found before
maximum during the 5 year mission lifetime is $\sim 8500$. This number
can be broken down into $\sim 6300$ SNe Ia, $\sim 500$ SNe Ib/c and
1700 SNe II (assuming $\f = 0.5$).  If data on a SN is taken before
maximum, then \GAIA\ has an excellent chance of identifying the
rapidly brightening object as a SN.

\subsection{Identification Strategy}

Every new object in a field of view is potentially a SN. Before \GAIA\
can identify a SN, it must have visited that location on the sky at
least once before. To provide SN alerts, we must ensure that the
brightening object is not just a common variable star.  We use the
General Catalogue of Variable Stars (Kholopov et al. 1999) to build a
subsample of variables with periods in excess of 10 days. Some $34 \%$
of this subsample have periods less than 6 months and $86 \%$ have
periods less than 1 year.  In practice, $\sim 12$ months baseline
photometry may be needed to discriminate against common forms of
stellar variability. One way round this problems is to restrict SN
alerts to high galactic latitudes ($|b| > 30^\circ$), where the
problem of variable star contamination is mitigated.

The objects that can cause most confusion are fast-moving solar system
asteroids and novae.  Main Belt asteroids move at $\sim 10$ mas
s$^{-1}$ and Near-Earth objects at $\sim 40$ mas s$^{-1}$ (e.g.,
Mignard 2002).  H{\o}g (2002) has shown that fast moving objects can
be detected by a single field of view crossing. This offers quick
discrimination between solar system asteroids and SNe.  More
problematical are novae.  Shafter (1997) gives the Galactic nova rate
as $35 \pm 11$ yr${}^{-1}$. We assume that this is typical of large
galaxies.  We take the absolute magnitude of novae to be in the range
$-6 < M < -9$ (Sterken \& Jaschek 1996). Given \GAIA's limiting
magnitude, there are between 20 and 150 galaxies in CfA catalogue for
which novae are detectable. This means that there are $\sim 1000$
novae per year in the \GAIA\ datastream. These can possibly be
distinguished from SNe on the basis of colour information and
spectroscopy. However, contamination by novae from external galaxies
-- which is the bulk of the numbers -- is restricted to a number of
small and pre-determined areas of the sky. These can, if necessary, be
excised from the SNe survey.

Therefore, a reasonable expectation is that \GAIA\ will alert on all
SNe caught before maximum, that is $\sim 1700$ SNe a year.  Riess et
al. (1998) show how the distance to a SNe Ia can be estimated to
within $10 \%$ from a single spectrum and photometric epoch.  For a
20th magnitude SN, the most suitable combination is a 2m telescope for
imaging and a 4m for spectroscopy. On a 2m telescope, it is feasible
to carry out high signal-to-noise UBVRI photometry on a 20th magnitude
SN in $\lta 1$ hr. The typical signal-to-noise needed for identifying
type from spectroscopy is $\sim 30$. This needs $\sim 2$ hrs on a $V
\sim 20$ point source in dark sky at low spectral resolution. However,
roughly half the SNe will not be suitable for follow-up from the
ground as they will be daytime objects. Assuming we wish to follow up
the sample of (nightime) SNe alerted before maximum, then roughly two
dedicated telescopes (say one 2m and one 4m) are required to get
distance and phase estimates. This will confirm detection and type.
Based on the information from the one night snapshot, selected SNe can
be chosen for more detailed monitoring. Candidates for intensive
monitoring might include all the SNe Ib/c and II-L (as there is little
information on their lightcurves), subluminous and over-luminous
events, all SNe Ia caught well before maximum and any SN for which the
snapshot gives an unusual luminosity or spectral composition. Tammann
\& Reindl (2002) have also recently emphasised the value for the
extragalactic distance scale of such a follow-up program of \GAIA\ SNe
Ia alerts.

\begin{table*}
\begin{center}
\begin{tabular}{l|ccc}\hline
SNe   & Total number & Fraction detected & Number detected \\
\null & detected & before maximum & before maximum \\
\hline
Type Ia   & $14308$     & $44 \%$ & $6317$\\
Type Ib/c & $1370$      & $37 \%$ & $ 501$\\
Type II-L & $8735 \f$   & $37 \%$ & $3205 \f$\\ 
Type II-P & $2683 (1-\f)$ & $9  \%$ & $236 (1-\f)$\\ 
Total     & $21387$     & $40\%$    & $8538$\\ \hline
\end{tabular}
\end{center}
\caption{The numbers of SNe of different types detected before maximum
during the 5 year \GAIA\ mission lifetime. The last row assumes that
the fraction of SNe II-L compared to all SNe II is $\f = 0.5$.}
\label{table:nosne}
\end{table*}
\begin{table*}
\begin{center}
\begin{tabular}{l|ccccccc}\hline
Distance  & Magnitude & SNe Ia & SNe Ib/c & SNe II-L & SNe II-P & Total \\
\hline
$< 50$ Mpc    & $G < 20$   & 23     & 16   &  $90 \f$   & $92 (1-\f)$  
& $\sim 130$ \\ 
$<100$  Mpc  & $G < 20$   & 179    & 107  &  $649 \f$  & $616 (1-\f)$ 
& $\sim 918$ \\
\hline
$< 50$ Mpc    & $G > 15$  & 8  &  10   &  $65 \f$   & $75 (1-\f)$  
& $\sim 88$ \\ 
$<100$  Mpc  & $G > 15$  & 115 & 79   &  $554 \f$  & $574 (1-\f)$ 
& $\sim 758$ \\
\hline
\end{tabular}
\end{center}
\caption{The numbers of SNe detected within 50 and 100 Mpc over 5
years. The upper table refers to all SNe brighter than 20th magnitude.
However, SNe brighter than 15th magnitude will probably be found
beforehand by other means. The lower table gives the numbers of all
SNe between 15th and 20th magnitude at maximum light. This gives an
idea of the numbers of nearby SNe that only \GAIA\ will find. The last
column assumes that the fraction of SNe II-L compared to all SNe II is
$\f = 0.5$.}
\label{table:near}
\end{table*}

\section{Scientific Returns}

\subsection{Follow-Ups, Neutrinos and Gravitational Waves}

Of course, it is important to follow up the alerted SN in bandwidths
other than the optical, including infrared, X-rays and gamma-rays.
This is needed -- amongst other things -- for the calculation of the
bolometric luminosity, which sums together all the flux associated
with the energy of the explosion and which provides tests and
constraints on the progenitor and detonation models (e.g., Mazzali et
al. 2001).  More exotically, the SN can also be sought with neutrino
and gravitational wave detectors, though these signals reach us before
the optical detection.

Infrared and optical photometry of a SN can be used to construct a
curve of colour versus time. By matching this to a template with zero
reddening, the extinction in the host galaxy can be measured (e.g.,
Krisciunas et al. 2001) and hence the absolute magnitude and distance
of the SN found.  There are additional benefits to infrared
follow-ups. One of the causes of uncertainty in SNe rates is
extinction.  For SNe Ib/c and II, which originate in the core collapse
of massive stars, the observed rates are likely to be a severe
underestimate of the true rate as extinction is usually high -- as
much as $10-20$ magnitudes -- in some starburst regions (Grossan et
al. 1999). Of course, the true SNe rate is important for understanding
chemical enrichment and evolution of galaxies and the ISM. Infrared
photometry, together with \GAIA\ data, will allow the rates to be
computed as a function of infrared (rather than blue) luminosity.

The X-ray and gamma-rays are associated with the radioactive decay of
nuclides produced in the SN explosion or Compton scattering associated
with high energy radiation. Only the explosions of SN 1987A and SN
1993J were detected in X-rays, whereas SN 1987A remains the only one
thus far detected in gamma-rays.  The present generation of
satellites, such as the {\it International Gamma-Ray Astrophysical
Laboratory} (INTEGRAL) can detect the gamma-ray signal of SNe out to
$\sim 10$ Mpc. However, as H\"oflich, Wheeler \& Khokhlov (1998) point
out, next generation Laue telescopes will allow gamma-rays from SNe
within $\sim 100$ Mpc to be detected. The numbers of SNe alerted by
\GAIA\ within 100 Mpc are listed in Table~\ref{table:near}. Some of
these SNe (typically those brighter than 15th magnitude) are likely to
have been found by other means beforehand. So, we conclude that \GAIA\
will alert on $\sim 920$ SNe in total for which the X-ray and
gamma-ray signal will be detectable. Of these, $\sim 760$ SNe will not
have been discovered by others beforehand.

Gravitational waves may also be sought from nearby SNe. For SNe Ib/c
and II, the asymmetry of the core collapse generates gravitational
waves which have a typical frequency $\sim 1$ kHz (e.g., M\"onchmeyer
et al. 1991). For SNe Ia, if the progenitor is a singly degenerate
binary (white dwarf and main sequence or red giant star), then no
detectable gravitational waves are expected. However, if the
progenitor is a double degenerate binary, then the collapse will look
like a SN II probably with a similar output of gravitational waves.
The maximal amplitude of gravitational waves for a SNe at 30 Mpc is in
the range $10^{-21}$ to $10^{-25}$ (e.g., M\"onchmeyer et al. 1991,
Bonnell \& Pringle 1995, Yamada \& Sato 1995).  The Laser
Interferometer Gravitational Wave Observatory (LIGO, see
``http://www.ligo.caltech.edu/'') is scheduled to begin taking data in
2003. The advanced LIGO will be operational after 2006 and can measure
gravitational waves with an amplitude down to $10^{-21}$. This raises
the possibility that gravitational waves from SNe within 10-50 Mpc may
be detected. From table~\ref{table:near}, there may be $\sim 130$ SNe
alerted by GAIA over 5 years for which detection of the gravitational
wave signal is feasible.  Note that gravitational waves arrive before
any photons.  However, the directional sensitivity of gravitational
wave detectors is very poor, so \GAIA\ may play an important r\^ole in
identifying the relevant SN.

Lastly, little is known about the neutrino emission from SNe Ia. For
SNe associated with core collapse, there are detailed theoretical
predictions.  However, the only SN for which neutrinos have been
detected remains SN 1987A.  Neutrinos were registered by the
Kamiokande II, the Irvine-Michigan-Brookhaven and the Baksan neutrino
detectors.  From the three detectors, a combined total of $\sim 25$
neutrinos with energies in the range 10-50 MeV are reckoned to
originate from SN 1987A (e.g., Burrows 1989).  These are believed to
be anti-electron neutrinos, for which the detection cross-section is
largest.  They were emitted in a burst of $\sim 10$ s associated with
the birth of the neutron star.  However, both global fits to the SN
parameters as well as theory suggest that the Kamiokande II yield was
smaller than expected, probably due to statistical fluctations and the
fact that the detector was not optimized at low energies. By
comparison, in the 32 kton Super-Kamiokande water-Cerenkov detector,
about $10^4$ detected events are expected from a core collapse SN at
an assumed distance of 10 kpc (Beacom, private communication). This is
larger than the SN 1987A yield in Kamiokande II by roughly the factors
25 (5 times closer), 16 (32 kton detector rather than 2 kton), and 2
(optimisation). In addition, some other reaction channels are
expected to become important once the yield is so high.

Scaling this result, the typical number of MeV neutrinos expected from
a SN Ib/c or II in a host galaxy at a distance $D$ with a
water-Cerenkov detector of volume $V$ is
\begin{equation}
N \sim 30\, \Bigl( {{\rm Mpc} \over D} \Bigr)^2 \Bigl( {V \over {\rm
Mton}} \Bigr).
\end{equation}
By \GAIA's launch date of 2010, it is reasonable to expect
1 Mton Hyper-Kamiokande detectors and so neutrinos from SNe at
distances up to $\sim 5$ Mpc may be detectable. This is the hardest
challenge of all!  Only a very few such nearby SNe are expected over
the 5 year \GAIA\ mission lifetime, and they may perhaps have been
found by other means before \GAIA\ alerts on them. However, even
today, very nearby SNe are still missed, particularly those that are
intrinsically faint or those occurring in obscured parts of the sky

\subsection{Applications}

Cosmological problems regarding the dark energy will not be addressed
directly by \GAIA. Even the most distant SNe that \GAIA\ detects have
$z \sim 0.14$.  By contrast, \GAIA\ will provide a large dataset of
nearby SNe. These are more interesting from the point of view of
understanding the properties and the underlying physics of the
explosions themselves.

The advantage of SNe surveys with \GAIA\ is that selection effects are
either minimised or easy to model, and that there will be many
examples of comparatively scarce phenomena (e.g., subluminous SNe, SNe
II-L, SNe Ib/c).  At present, SNe rates come from relatively small
datasets (e.g., Hardin et al. 2000) and are subject to substantial
uncertainties. Selection effects -- depending on the type of host
galaxy, the extinction and the distance from the center of the galaxy
-- seriously afflict all current datasets.  Given the large numbers of
alerted SNe, \GAIA\ will provide accurate rates as a function of
position, extinction and type of host galaxy.  These give valuable, if
indirect, information on both the star formation rate and the high
mass end of the mass function. There have also been suggestions that
populations of subluminous SNe may have been systematically missed in
existing catalogues (e.g., Richardson et al. 2002). If so, then \GAIA\
is the ideal instrument with which to find them. The value of \GAIA\
in richly populating the Hubble diagram has also been pointed out
recently by Tammann \& Reindl (2002).

The nature of the progenitor populations of the different types of SNe
-- and especially SNe Ia's -- will probably remain unsolved over the
next decade.  \GAIA\ can help as it provides milliarcsecond astrometry
or better with a single transit.  For a SN at a typical distance of
$\sim 200$ Mpc, the projected position can be determined to $\sim 1$
pc (assuming the main error contribution comes from angle
measurements). There are only two detections of a SN progenitor before
explosion (namely SN 1987A and SN 1993J). Attempts have been made to
locate the progenitors of nearby SNe using astrometry with an accuracy
of 0.17'', though without success so far (Smartt et al. 2002).  \GAIA\
will provide milliarcsecond error boxes and therefore it may be easier
to locate the progenitor for nearby SNe using HST or other deep search
archives. Even if no star is visible, then constraints can still be
placed on the progenitor mass, as Smartt et al. (2002) demonstrate.
Accurate positional information on the location of SNe within galaxies
is important because it sheds light on the spatial distribution (and
hence nature) of the progenitor population.

Zampieri et al. (1998) and Balberg \& Shapiro (2001) emphasise that
the formation of the black hole may also be detectable in nearby SNe
as luminosity generated by late-time accretion. This is only possible
in the case of subluminous SNe, as otherwise it is masked by radiative
heating.  Using Figure 4 of Balberg \& Shapiro (2001), we see that the
fraction of suitable SNe comprise roughly $10 \%$ of all core collapse
SNe within 50 Mpc. From Table~\ref{table:near}, it follows that there
will be $\sim 10$ SNe Ib/c and II with the required properties alerted
by \GAIA\ during the whole mission. If subluminous SNe have been
systematically missed in current catalogues, then this number will be
still higher.  For these SNe, follow-ups may be able to detect the
faint luminosity indicative of black hole accretion. This requires
scheduling observations with instruments like the {\it Hubble Space
Telescope} or the {\it Next Generation Space Telescope} of the
declining light curve with the hope of observing accretion onto the
emerging black hole.

Another application of \GAIA's SNe Ia dataset is to measurements of
the local velocity field (e.g., H{\o}g et al 1999, Tamman \& Reindl
2002). Nowadays, the bulk flow and the bias parameter are often
calculated using distance estimators of galaxies such as the
Tully-Fisher relationship or the fundamental plane, combined with
radial velocity measurements (e.g., Hudson et al. 1999). Typically,
this proceeds by finding the best fitting bulk flow ${\bf V}$ which
minimises
\begin{equation}
\chi^2 = \sum_i {(v_i - {\bf V}.{\bf {\hat r}_i})^2 \over \sigma_i^2},
\end{equation}
where $v_i$ is the observed peculiar velocity for the ith galaxy,
whose direction vector is ${\bf {\hat r}_i}$, and $\sigma_i$ is an
estimate of the error.  As pointed out by Riess et al. (1997), SNe Ia
offer a more accurate distance estimator than galaxies, as the typical
uncertainty is reduced to $\sim 5 \%$. In other words, every SNe Ia
identified by \GAIA\ can give the distance of its host galaxy with
unprecedented accuracy.  When combined with ground-based spectroscopy,
\GAIA\ will provide $\sim 6300$ positions and peculiar velocities of
galaxies in the nearby Universe.  These can be used to study the
deviations from the Hubble flow and compared with the velocities
predicted by gravity fields of full-sky galaxy catalogues. With such
data, not merely the local bulk flow but also the shear field will be
measurable.

\section{Conclusions}

The theory of SNe explosions is well-developed but somewhat poorly
calibrated against data.  Nearby SNe are ideal for carrying out such
detailed comparisons. Provided we can alert on a large enough sample
of SNe, then there are good prospects for the identification of the
progenitor population and the detection of the ongoing explosion in
all wavebands. All these provide tests and checks on the theory of
stellar evolution and SNe detonation.  The identification of the
gamma-ray and gravitational wave signals may also be possible for
close SNe (within 50 Mpc).

We have demonstrated that the astrometric scanning satellite \GAIA\
has exactly the required capabilities.  \GAIA\ is the European Space
Agency satellite that is the successor to \Hipparcos. Over \GAIA's
five year mission lifetime, we estimate that the numbers caught before
maximum are $\sim 6300$ SNe Ia, $\sim 500$ SNe Ib/c and $\sim 1700$
SNe II. After launch in about 2010, \GAIA\ will issue $\sim 5$ SN
alerts a day. In total, \GAIA\ will report data on at least $21\,400$
SNe during the mission, of which $\sim 14\,300$ are SNe 1a, $\sim
1400$ are SNe Ib/c and $\sim 5700$ are SNe II.  These numbers come
from detailed simulations in which SNe explode in galaxies tracing the
local large-scale structure. The effects of the dimming of SNe because
of absorption in the host galaxy and in the Milky Way are
included. The SNe lightcurves are sampled according the scanning law
of the satellite.  We note that, though huge, these numbers are lower
limits. They do not take into account the contribution of SNe in
low-luminosity galaxies, which may result in a further doubling of the
numbers. They do not take into account the fact that \GAIA's limiting
magnitude is likely to be somewhat deeper than 20th in practice.

\GAIA\ is complementary to the {\it SNAP} satellite (see
``http://snap.lbl.gov''), which will probe much deeper out to a
redshift of $\sim 1.7$ to obtain $\sim 2000$ SNe Ia a year. By
contrast, \GAIA\ will obtain data on $\sim 4280$ SNe a year but probe
at most out to a redshift of $\sim 0.14$. \GAIA\ will play its role in
cosmology by providing a huge dataset of high quality local SNe Ia in
which any scatter in the Phillips relation with colour can be
quantified and analysed (Tammann \& Reindl 2002).

\GAIA\ will provide an important database of nearby SNe which are
particularly interesting for studies of the explosions
themselves. Many statistics -- SNe rates, frequency of different
lightcurve morphologies, location in the host galaxy -- are poorly
known.  \GAIA\ will provide the definitive dataset. In particular, SNe
rates can be corrected for known biases, such as extinction and
position in the host galaxy, and so will be measured with
unprecedented accuracy. The huge numbers mean that even examples of
comparatively rare phenomena will be present in the database.  Despite
the enormous efforts of the last decade, there are still very few
subluminous SNe, over-luminous SNe, SNe II-L and SNe Ib/c in the
standard catalogues. \GAIA\ will present us with good chances of
finding examples of nearby subluminous SN Ib/c and II, which needed to
detect the emergence of the black hole.  Every SNe Ia identified by
\GAIA\ can give the distance of its host galaxy with better accuracy
than any other method. Thus, the density and velocity field within a
few hundred Mpc will be delineated with unprecedented precision.

\section*{Acknowledgments}
We wish to thank James Binney, Erik H{\o}g and Michael Perryman for
encouragement and information about the \GAIA\
satellite. Additionally, John Beacom, Bruno Leibundgut, Philipp
Podsiadlowski, Serge Popov, Steve Smartt and Chris Tout provided
helpful advice on supernovae. L. Lindegren and R. Drimmel kindly
supplied software on the scanning law and on the Galactic extinction
model respectively.  VB acknowledges support from the Dulverton Fund,
while NWE is supported by the Royal Society. Additionally, we thank
the anonymous referee for some helpful criticisms that improved the
paper.

%REFERENCES
{}

\end{document}